\documentclass[twocolumn,showpacs,superscriptaddress]{revtex4}
\usepackage{graphicx}
\usepackage{natbib}
\usepackage{latexsym}
\begin{document}
%\draft\twocolumn[\hsize\textwidth\columnwidth\hsize\csname
%@twocolumnfalse\endcsname
\title{Atom-Atom Scattering Under Cylindrical Harmonic
Confinement: Numerical and Analytic Studies of the Confinement Induced Resonance}
\author{T. Bergeman}
\email{thbergeman@notes.cc.sunysb.edu} \affiliation{Permanent Address: Department of Physics
and Astronomy, SUNY, Stony Brook, NY 11794-3800} \affiliation{ITAMP, Harvard, Cambridge,
Massachusetts 02138}
\author{Michael Moore}
\email{mmoore@cfa.harvard.edu} \affiliation{ITAMP, Harvard, Cambridge, Massachusetts 02138}
\author{M. Olshanii}
\email{olshanii@phys4adm.usc.edu} \affiliation{Permanent address: Physics Department,
University of Southern California, Los Angeles, CA} \affiliation{ITAMP, Harvard, Cambridge,
Massachusetts 02138}
\date{\today }
\begin{abstract}
M. Olshanii [Phys. Rev. Lett. {\bf 81}, 938 (1998)] recently solved the atom-atom scattering
problem with a pseudopotential interaction in the presence of transverse harmonic
confinement, i.e. within an `atom waveguide', deriving an effective one-dimensional coupling
constant that diverged at a ``confinement induced resonance'' (CIR). Here, we report
numerical results for finite range potentials that corroborate this resonance. In addition,
we now present a physical interpretation of this effect as a novel type of Feshbach
resonance in which the transverse modes of the waveguide assume the roles of `open' and
`closed' scattering channels.
\end{abstract}
\pacs{03.75Fi,05.30.Jp,05.10.-a}
\maketitle

Recently, there have been a number of experiments in which ultracold atoms and/or
Bose-Einstein condensates have been loaded into magnetic or optical `atom waveguides'
\cite{Bongs,Hansch,Denschlag,Hinds,Prentiss,Greiner,Arlt,DWeiss}. One goal of such
experiments is to reach the `single-mode' or quasi-1D regime, where only the ground state of
transverse motion is significantly populated at thermal equilibrium. This regime is of great
practical interest due to the potential  for ultra-sensitive rotation and gravitational
gradient detection with guided single-mode atom interferometers. In addition to such
applications, reaching the quasi-1D regime is of significant theoretical interest as the 1D
delta-interacting boson gas represents one of the few known fully integrable quantum field
theories. In a finite system with infinitely strong (hard-core) delta-function interactions,
boson many-body states in 1D have been shown to correspond via a one-to-one mapping with the
highly-correlated states of the corresponding non-interacting Fermi gas \cite{Girardeau}.
The properties of this Tonks-Girardeau gas have been a topic of significant current
theoretical interest \cite{Tonks,Girardeau,GWT,DLO,Petrov,Stoof,DGW} in anticipation of
future atom-waveguide experiments. Additionally, the homogeneous 1D Bose with
arbitrary-strength delta-function interactions, known as the Lieb Liniger model, is also a
fully integrable system \cite{LL}.

To make the connection between experiments in tightly confining waveguides and theoretical
models in 1D, it is necessary to know the relationship between the effective 1D coupling
constant, $g_{1D}$, and the 3D scattering length, $a$. This problem was first addressed
rigorously in \cite{MOPRL}, where it was predicted that a `confinement induced resonance'
(CIR) modifies the effective interaction, resulting in an effective 1D coupling strength
which can be tuned from $-\infty$ to $+\infty$ by varying the transverse width of the
waveguide, $a_\perp$ over a small range in the vicinity of the resonance at $a_\perp=Ca$,
where $C=1.4603\ldots$. Hence this resonance clearly has significant implications for
atom-waveguide experiments. Until now, however, there has been no convincing physical
explanation for the effect, thus raising questions concerning its appearance in systems with
finite-range interactions.

The primary goal of this Letter, therefore, is to present numerical calculations of
scattering in the presence of a cylindrical harmonic potential using finite-range atom-atom
potentials, confirming the existence of the CIR. In addition, we provide a much-needed
physical interpretation of the effect as a Feshbach-type resonance involving bound states of
the energetically closed transverse modes (`channels'). Analogous numerical scattering
studies in spherically symmetric traps have been reported \cite{Tiesingapp,Bolda}, while
scattering in the presence of harmonic confinement in one dimension is discussed in
\cite{PHS,PS}.

We begin our analysis by considering a collision between two atoms initially in the ground
state of transverse motion. In the presence of harmonic transverse confinement the
center-of-mass motion and relative motions are separable, with the wavefunction of the
relative coordinate satisfying an effective single-particle model with a stationary
scatterer at the origin. Furthermore, if the longitudinal kinetic energy in the
center-of-mass frame is less than the transverse level spacing then the atoms remain
asymptotically frozen in the ground state. Low-energy scattering in this regime can then be
modeled in the pseudopotential approximation \cite{Demkov} by the 1D Hamiltonian
\begin{equation} \label{H1D}
    H_{1D}=-\frac{\hbar^{2}}{2\mu} \frac{\partial^{2}}{\partial z^{2}} +
    g_{1D} \delta(z),
\end{equation}
where $z$ is the longitudinal atomic separation and $\mu$ is the reduced mass. In the
pseudopotential approximation, $g_{1D}$ is typically obtained as in \cite{DGW} by assuming
that the wavefunction of the relative coordinate ${\bf r}={\bf r}_1-{\bf r}_2=
z\hat{z}+\rho\hat{\rho}$ factorizes as $\Psi({\bf r}) =\phi_0(\rho)\psi(z)$, where
$\phi_{0}(\rho) = \exp(-\rho^{2}/2a_{\perp}^{2})/(a_{\perp}\sqrt{\pi})$ is the transverse
ground state, and $a_{\perp} = \sqrt{\hbar/\mu \omega_{\perp}}$ is the transverse harmonic
oscillator length for the relative atomic motion, $\omega_\perp$ being the transverse trap
frequency. The effective 1D potential is then defined via $\int 2\pi\rho d\rho
|\phi_0(\rho)|^2\frac{2\pi\hbar^2a}{\mu}\delta^3({\bf r})=g_{1D}\delta(z)$, where $a$ is the
3D scattering length, which leads to
\begin{equation} \label{g1Dapprox}
    g_{1D}\approx \int_{0}^{\infty}2 \pi \rho d\rho
    =\frac{2\hbar^{2} a}{\mu a_{\perp}^{2}}.
\end{equation}

From an exact solution of the three-dimensional scattering problem with a zero-range s-wave
interaction, however, it was recently predicted \cite{MOPRL} that a ``confinement induced
resonance'' (CIR) modifies the effective interaction. By matching the low-energy scattering
amplitude of the exact solution to that of Eq. (\ref{H1D}), it was found that:
\begin{equation} \label{g1D}
    g_{1D} = \frac{2 \hbar^{2} a}{\mu a_{\perp}^{2}}
\frac{1}{(1 -C a/a_{\perp})},
\end{equation}
where $C =-\zeta(1/2)= 1.4603..$. We note that the use of Eqs. (\ref{H1D}) and (\ref{g1D})
requires $ka_\perp\ll 1$, where $\hbar k$ is the collision momentum. The appearance of the
resonance term in the denominator implies that to obtain an infinite delta-function
interaction (thereby accessing the Tonks-Girardeau regime), it is sufficient to satisfy the
CIR condition $a_\perp \approx Ca$. From the naive formulation (\ref{g1Dapprox}) one would
conclude that accessing this regime requires the more extreme condition $a\gg n a_\perp^2$,
where $n$ is the linear density.

Before addressing the physical interpretation of the CIR, we first
describe our numerical results pertaining to low-energy scattering
and the effective 1D coupling constant with finite-range
interactions. The relation between $g_{1D}$ and the 1D scattering
amplitude may be found by assuming that the scattering eigenstates
$\Psi(z)$ of Hamiltonian (\ref{H1D}) take the form given in
\cite{MOPRL}
\begin{equation} \label{Psi1D}
    \Psi(z) = e^{ikz} + f_{e} e^{ik|z|}; \ \ E = \hbar^{2} k^{2}/2 \mu,
\label{scatt._solution}
\end{equation}
where $f_{e}$ is the coefficient for the even part of $\Psi$, while the odd part can be
shown to vanish by continuity arguments. From Eq. (\ref{Psi1D}), we find $\Psi''(z) = -
k^{2} \Psi(z) + 2ik f_{e}(k) \delta(z)$, which must equal $(2 \mu/\hbar^{2}) g_{1D} \Psi(z)$
in the limit $k\to 0$. This gives
\begin{equation} \label{g1Dfe}
    g_{1D} = \lim_{k\to 0}\frac{\hbar^{2}kf_{eR}(k)}{\mu f_{eI}(k)}
\end{equation}
where $f_{eR}$ and $f_{eI}$ are the real and imaginary parts of
$f_{e}$. To extract analogous values for $g_{1D}$ from numerical
scattering calculations, we obtain eigenfunctions of the
Hamiltonian $\hat{H}=\hat{H}_z + \hat{H}_\perp+\hat{V}$, where
\begin{eqnarray}  \label{Ham}
    \hat{H}_z =-\frac{\hbar^{2}}{2\mu}
    \frac{\partial^{2}}{\partial z^{2}}; \
    \hat{H}_{\perp}=-\frac{\hbar^{2}}{2 \mu}
    \left[ \frac{\partial^{2}}{\partial \rho^{2}}+\frac{1}{\rho}
    \frac{\partial}{\partial \rho} \right]+
    \frac{\mu}{2}\omega_{\perp}^{2} \rho^{2} .
\end{eqnarray}
In the present work we restrict ourselves to the case of zero azimuthal angular momentum
$m$=0, as there is negligible s-wave scattering for $m \neq 0$. For the atom-atom potential,
we will study two cases: $V(r) = C_{12}/r^{12}- C_{6}/r^{6} (r^{2} = z^{2} + \rho^{2})$ and
the spherical square-well, $V(r) = -\bar{V}S(b-r)$, $S(r)$ being the unit step function.

With application to the case of Cs atoms in a 1D optical well \cite{DWeiss} in mind, we
consider atoms with the mass of $^{133}$Cs, and with the $C_{6}$ coefficients as determined
recently \cite{Leo} to be $C_{6}$(Cs) = 6890 a.u. For Cs atoms in the $(F,M)$ = (3,3) state,
the scattering length is determined only as an upper bound, $a(Cs) <$ -140 nm, as compared
with $a_{\perp}$ = 29.5 nm reported in\ cite{DWeiss}. In order to study a more general
situation, we allow $C_{12}$ to vary.  To simplify the numerics we consider the regime of
just 1 to 3 $J=0$ bound states, rather than the 47 bound states in the actual Cs$_{2}$
$^{3}\Sigma_{u}^{+}$ state, and we neglect other terms in the dispersion potential. The 6-12
potential may be characterized by $R_{e}$, the minimum of the potential well, where $R_{e} =
(2 C_{12}/C_{6})^{1/6}$.  For $C_{6} = C_{6}$(Cs) and for a series of values of $C_{12}$ we
obtain the free-space $s$-wave scattering length, $a$, by Numerov integration of the
Schr\"{o}dinger equation.  When a resonance state passes through threshold $a$ exhibits a
simple pole, thus the 6-12 potential provides the full range of 3D scattering lengths. For
the second case of a spherical well potential the scattering length for a well of depth
$\bar{V}$ and range $b$ is given by $a = b - \tan(b \eta)/\eta$, where $\eta^{2} = 2 \mu
\bar{V}/\hbar^2$.

To solve the scattering problem for such central potentials plus a transverse harmonic
potential, we employed a numerical mesh in $\rho$ and $z$, and found eigenfunctions in a
cylindrical box of finite length. The box was sufficiently long in $z$ that the asymptotic
form of the wavefunction as $|z|\to\infty$ could be determined. The eigenfunctions were of
odd or even parity, with the odd parity functions exhibiting negligible scattering effects.
For even functions of energy $E = \hbar \omega_{\perp} + \hbar^{2}k^{2}/2 \mu$, the
asymptotic form is
\begin{equation}
\label{FG}
    \Psi(\rho,z) \stackrel{|z| \rightarrow \infty}{\longrightarrow}
    {\cal N} \left[(1 + f_{eR})\cos{(kz)} - f_{eI} \sin{(k|z|)}\right]\phi_{0}(\rho).
\end{equation}
Values for $f_{eR}$, $f_{eI}$ and ${\cal N}$ were extracted from the coefficients of
$\sin{(kz)}$ and $\cos{(kz)}$ in Eq. (\ref{FG}), determined over a range of $z$ values for
which $V(r)$ is negligible, and from conservation of probability current. From $f_{eR}$ and
$f_{eI}$, $g_{1D}$ is then obtained from Eq. (\ref{g1Dfe}) via extrapolation of finite $k$
data to $k=0$. The numerical mesh was provided by the discrete variable representation
(DVR). In order to increase the density of points near $z$=0, the $z$ coordinate was scaled
by $z=U(y)$, where $U(y) = a \cosh(z/b)$, analogous to scaling used in \cite{Tiesinga}.  A
uniform mesh in $y$ was then used \cite{Colbert} along with a Laguerre DVR in $\rho$
\cite{BayeHeenan}. Eigenfunctions in first iteration were found by exact diagonalization
over a relatively small mesh ($<$10,000 mesh points).  As additional mesh points were added,
sparse matrix diagonalization techniques were used.
\begin{figure}
\vspace*{-10mm}
\includegraphics[scale=.9]{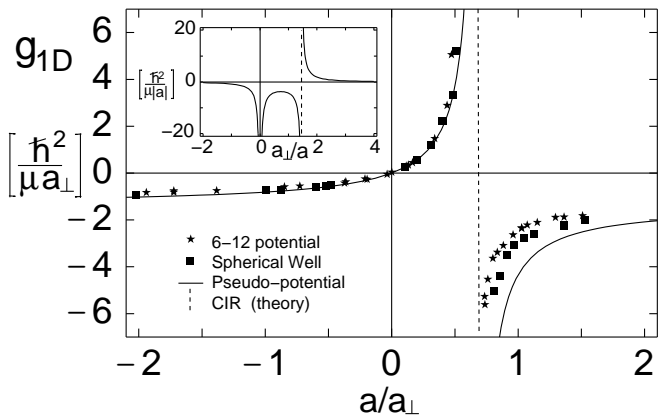}
\caption{\label{g1Dfig}The 1D coupling constant, $g_{1D}$, in units of $\hbar^2/(\mu
a_\perp)$, as a function of $a/a_{\perp}$ for the 6-12 potential (stars) and the spherical
well potential (triangles), as compared to the pseudopotential theory (solid line). The
inset shows $g_{1D}$ in units of $\hbar^2/(\mu |a|)$ versus $a_\perp/a$, to illustrate the
behavior in the tight-confinement regime ($a_\perp/a\to 0$).}
\end{figure}

Numerical results for $g_{1D}$ with the 6-12 potential and with the spherical well are shown
in Fig. \ref{g1Dfig}, in comparison with the analytic result (\ref{g1D}).  The numerical
calculations clearly exhibit a singularity in $g_{1D}$ at $a/a_\perp = 1/C$. For $a$ less
than this value the calculated $g_{1D}$ values agree well with the analytic expression.
There do appear to be systematic deviations between the exact numerical results and the
pseudopotential theory that are greater for the 6-12 potential than for the spherical well
for $a/a_\perp>1/C$. This effect is connected to the Feshbach resonance interpretation and
occurs because the bound state of the pseudopotential does not agree precisely with the
upper bound-state of the finite-range potentials having the same scattering length $a$.

These numerical results validate the pseudopotential analytic result \cite{MOPRL},
demonstrating that the CIR is indeed a physical phenomena rather than an artifact of the
pseudopotential approximation. It is therefore important to establish the physical origin of
the CIR. In what follows we are going to show that the CIR is in fact a zero-energy Feshbach
resonance, occurring when the energy of a bound state of the asymptotically closed channels
(i.e. the excited transverse modes) coincides with the continuum threshold of the open
channel (lowest transverse mode).

To illustrate this interpretation we consider the total Hamiltonian $\hat{H}=\hat{H}_{z} +
\hat{H}_{\perp}+\hat{V}_{pseudo}$, where the pseudopotential operator is defined by
$\langle{\bf r}|\hat{V}_{pseudo}|\psi\rangle=\frac{2\pi\hbar^2 a}{\mu}\delta^3({\bf
r})\frac{\partial}{\partial r}r\langle{\bf r}|\psi\rangle$. Because our interpretation of
the CIR relies heavily on the behavior of two-atom bound states, it is important to recall
that the Huang-Fermi pseudopotential, $\hat{V}_{pseudo}$, supports a single bound state in
free space with energy $\hbar^2/(\mu a^2)$, provided $a>0$. With the existence of this bound
state in mind, we proceed by formally splitting the Hamiltonian onto `ground', `excited',
and `ground-excited coupling' parts according to
\begin{eqnarray}
\label{manifolds}
    \hat{H}&=& \hat{H}_{g}+\hat{H}_{e}+\hat{H}_{g-e}\nonumber\\
    &=&\hat{P}_{g}\hat{H} \hat{P}_{g}+\hat{P}_{e}\hat{H}\hat{P}_{e}
    +(\hat{P}_{e}\hat{H}\hat{P}_{g}+\mbox{h.c.}),
\end{eqnarray}
where $\hat{P}_{g} = |0\rangle \langle 0|$, $\hat{P}_{e} = \sum_{n=1}^{\infty} |n\rangle
\langle n|$, are the corresponding projection operators, $|n\rangle$ being the eigenstate of
the transverse two-dimensional harmonic oscillator with {\it radial} quantum number $n$ and
zero axial angular momentum (we recall that in defining $\hat{H}_\perp$ we have already
projected out the states with non-zero axial angular momentum). The corresponding
eigenvalues of the transverse Hamiltonian are given by $\hat{H}_{\perp} |n,\rangle =
\hbar\omega_{\perp} (2n +1) |n\rangle$.

The `ground' Hamiltonian has a 1-d coordinate representation of the form of Eq. (\ref{H1D}),
corresponding to the motion of a one-dimensional particle in presence of a $\delta$-barrier
with the `bare' coupling constant given by Eq. (\ref{g1Dapprox}). The spectrum of
$\hat{H}_g$ is continuous for energies above the threshold energy $E_{C,g} =
\hbar\omega_{\perp}$ (${\cal E}=0$) and contains a single bound state at energy
$E_{B,g}<E_{C,g}$ for the case of positive 3D scattering length. Likewise, the spectrum of
the `excited' Hamiltonian is clearly continuous for energies $E_{C,\,e} =
3\hbar\omega_{\perp}$ (${\cal E} = 1$), but as we will see below, $\hat{H}_{e}$ supports one
bound state of an energy $E_{B,e}<E_{C,e}$ for {\it all} values of the 3D scattering length,
$a$.

It is clear that the remaining coupling between the `ground' and `excited' subspaces will
renormalize the bare coupling (\ref{g1Dapprox}). Furthermore, according to the Feshbach
scheme, one would predict a resonance in the renormalized $g_{1D}$ for a set of parameters,
such that the energy of the bound state of $\hat{H}_{e}$ coincides with the continuum
threshold of $\hat{H}_{g}$:
\begin{eqnarray}
    E_{B,e} = E_{C,g} \quad \Rightarrow \quad \mbox{CIR}
\label{CIR_condition}
\end{eqnarray}
As we will see below this scheme indeed predicts a position of the
CIR {\it exactly}.

The energy $E_{B,\,e}$ of the bound state of $\hat{H}_{e}$ can be
found using the following two step procedure. First we identify
the bound state energy of the full Hamiltonian $\hat{H}$ as a pole
of the scattering amplitude on the physical Riemann sheet. Second,
we make use of the peculiar property of the two-dimensional
harmonic oscillator that `excited' Hamiltonian $\hat{H}_{e}$and
the full Hamiltonian $\hat{H}$ can be transformed to each other
via a simple unitary transformation. This allows for simple
relation between their energy spectra, in particular the energy of
the bound states.

The even-wave one-dimensional scattering amplitude $f_{e}$ at an
energy $E$ ($E_{c,\,g} \le E < E_{c,\,e}$), as defined in
(\ref{scatt._solution}), has been derived in \cite{MOPRL}. We have
now obtained a closed-form analytic expression for the previously
derived result, so that the scattering amplitude can be expressed
as
\begin{equation}
    f_{e}(k) = -\, \frac{2i}{ka_{\perp}\lbrack \frac{a_{\perp}}{a}
    +\zeta(1/2,\,-(ka_{\perp}/2)^2) \rbrack}
\end{equation}
where
\begin{eqnarray}
\label{Hurwitz}
    &&\zeta(s,\,\alpha) = \lim_{N\to\infty}\, \sum_{n=0}^{N}
    \frac{1}{(n+\alpha)^{s}} - \frac{(N+\alpha)^{-s+1}}{-s+1};\nonumber\\
    &&Re(s) > 0, \,\, Im(s) = 0,\nonumber\\
    &&z^{s} = |z|^{s}\,e^{is (\mbox{\small Arg}(z)-2\pi)}, \,\, 0 < \mbox{Arg}(z) \le 2\pi
\end{eqnarray}
is the Hurwitz Zeta function \cite{Apostol}, and the wave-vector
$k$ is given by $E = E_{C,g} + \hbar^2 k^2/2\mu$. The bound state
energies of the full Hamiltonian $\hat{H}$ will be given by the
poles, $\bar{k}$, on the {\it positive imaginary} axis of the
analytic continuation of $f_e(k)$: $E_{B,full} = -\hbar^2
Im^2(\bar{k})/2\mu$. One can see that in order to avoid crossing
the branch-cuts of the Zeta function, the continuation should be
performed inside the $0 \le \mbox{Arg}(k) \le \pi/2$ quadrant of
the complex plane. We find a single pole corresponding to the
following implicit equation for the bound state energy:
$\zeta(1/2,\, -E_{B,\,full}/(2\hbar\omega_{\perp}) + 1/2) =
-a_{\perp}/a$.

Notice now, that the full Hamiltonian $\hat{H}$ and the excited Hamiltonian $\hat{H}_{e}$
are connected via a simple transformation: $\hat{H}_{e} = \hat{A}^{\dagger}\hat{H}\hat{A}$,
where $\hat{A}^{\dagger} = \sum_{n=0}^{\infty} |n+1,\rangle \langle n|$. Note that both
$\hat{H}_{e}$ and $\hat{H}$ include interactions, thus the above property is highly
nontrivial and stems from the fact that the $m=0$ eigenfunctions of the two-dimensional
harmonic oscillator all have the same value, $1/\sqrt{\pi}a_{\perp}$, at the origin. Thus
the three dimensional $\delta$-interaction has {\it the same} matrix elements between all
the harmonic oscillator states so that interaction matrix is unaffected by the shift
operator.
\begin{figure}
\vspace*{-10mm}
\includegraphics[scale=.75]{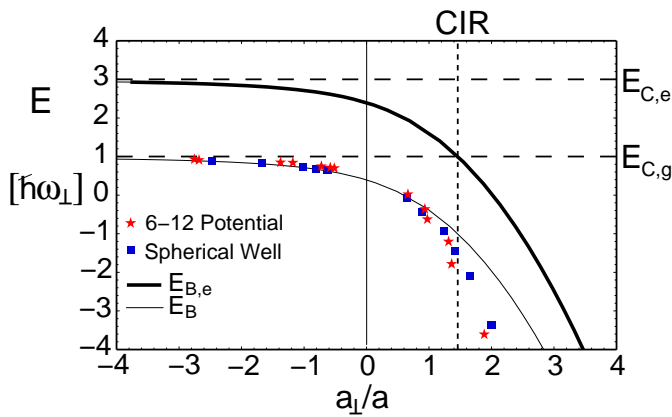}
\caption{\label{feshbach.fig}Numerical bound state energies and schematic of the Feshbach
resonance scheme. The solid lines correspond to the analytic pseudopotential results.}
\end{figure}

From the above we conclude that the bound state energy $E_{B,e}$ of the `excited'
Hamiltonian is related to the bound state energy $E_B$ of the full Hamiltonian via $E_{B,e}
= E_B + 2\hbar\omega_{\perp}$, and thus obeys an implicit equation
$\zeta(1/2,-E_{B,e}/(2\hbar\omega_{\perp}) + 3/2) = -a_{\perp}/a$. The CIR condition
(\ref{CIR_condition}) can be now explicitly formulated as $\zeta(1/2,
-E_{C,g}/(2\hbar\omega_{\perp}) + 3/2) = -a_{\perp}/a$. Using $E_{C,g} =
\hbar\omega_{\perp}$, we finally arrive at the exact CIR condition
$\frac{a_\perp}{a}=\zeta(1/2,1)=\zeta(1/2,0)=-C$. A similar effect is associated with
resonance behavior in harmonically confined 2D scattering for $a<0$ \cite{PHS}. We note that
this resonance would most-likely be observed via changes in the macroscopic properties of
the ground-state of a many-atom system, i.e. density distribution, as described in
\cite{DLO}.

This Feshbach scheme is illustrated in Figure 2, where we plot the bound-state energies
$E_{B,e}$ (dark solid line) and $E_B$ (thin solid line) as a function of the ratio
$a_\perp/a$. The continuum thresholds $E_{C,e}$ and $E_{C,g}$ are also indicated,
illustrating that the CIR occurs when the bound state of the manifold of closed channels,
$E_{B,e}$, crosses the continuum threshold of the open channel, $E_{C,g}$. In addition, we
have plotted the bound state energies of the full Hamiltonian as determined numerically for
the $6-12$ and spherical well potentials, showing good agreement with the pseudopotential
result. As the bound-state energy deepens, we start to see quantitative disagreement between
the bound state energies of the finite range potentials and the pseudopotential. This
disagreement is consistent with the discrepancy in the position of the CIR shown in Fig. 1,
showing that it is the bound state energy, and not the scattering length which determines
the location of the CIR. Lastly, we note that while in free space a weakly bound state
exists only for $a>0$, we see that in the waveguide such a state exists for all $a$. These
bound states may be of significant interest, allowing the formation of dimers via a
modulation of the waveguide potential at the frequency $(E_{C,g}-E_B)/\hbar$. This may lead
to an atom-waveguide based scheme for forming ultracold dimer molecules, as well as the
possibility to use molecular spectroscopy as a sensitive probe of the atomic field inside
the waveguide.

This work was supported by the National Science Foundation through
a grant for the Institute for Theoretical Atomic and Molecular
Physics at Harvard University and Smithsonian Astrophysical
Observatory. The work at Stony Brook was supported by the Office
of Naval Research, and the work at USC by the NSF grant {\it
PHY-0070333}.  We thank D. S. Weiss for discussions that provided
part of the motivation for this study and R. Shakeshaft for
discussions on Feshbach resonances.

\end{document}